\documentclass[preprint]{revtex4-1}

\usepackage{bm}
\usepackage{amsmath,amssymb}
\usepackage{graphicx,subfig}
\usepackage{psfrag} 
\usepackage{rotating}
\usepackage{color}

\begin{document}

\title{Simulating the interaction between a falling super-quadric object and a soap film}

\author{I. T. Davies} 
\email{itd@aber.ac.uk}
\affiliation{Department of Mathematics, Institute of Mathematics, Physics and Computer Science, Aberystwyth University, Aberystwyth, Ceredigion SY23 3BZ, UK} 
\affiliation{Coleg Cymraeg Cenedlaethol, Y Llwyfan, Heol y Coleg, Caerfyrddin, SA31 3EQ, UK}

\begin{abstract}
The interaction that occurs between a light solid object and a horizontal soap film of a bamboo foam contained in a cylindrical tube is simulated in 3D. We vary the shape of the falling object from a sphere to a cube by changing a single shape parameter as well as varying the initial orientation and position of the object. We investigate in detail how the soap film deforms in all these cases, and determine the network and pressure forces that a foam exerts on a falling object, due to surface tension and bubble pressure respectively. We show that a cubic particle in a particular orientation experiences the largest drag force, and that this orientation is also the most likely outcome of dropping a cube from an arbitrary orientation through a bamboo foam.   
\end{abstract}

\maketitle

\section{Introduction}
	\label{sec:Introduction}

Liquid foams are a class of materials that are widely-used both domestically and industrially \cite{FrenchBook,Weaire2000}. They are classified as complex fluids because their response to applied stress is highly non-linear; they behave as elastic solids at low stresses, exhibit plasticity at higher stresses and have an apparent yield stress above which they flow. They are two-phase materials consisting mainly of gas, therefore have a low density but also a large surface area. As a result of these remarkable properties, aqueous foams are desirable for example in personal hygiene and food products and are also integral in industrial processes such as froth flotation \cite{Prudhomme96,Shean11} and enhanced oil-recovery \cite{Rossen96,Sheng13}. Recently, foams have been studied at the microfluidic scale \cite{Drenckhan05,Raven06} where they have found new applications in medical procedures such as foam sclerotherapy for spider and varicose veins and for building new materials such as scaffolds for tissue engineering \cite{Huerre14,Whitesides06}.

In this work, we consider another possible application for liquid foams and ordered aqueous foams in particular, where their precise structure could be used to control the position and orientation of particles or small objects. We focus our attention on the interaction that occurs between a single soap film that separates two bubbles in an ordered bamboo foam contained in a cylinder and a solid object that falls through it under gravity. We develop 3D Surface Evolver \cite{Brakke} simulations to investigate how this interaction affects the final position and orientation of a non-spherical falling object. This work follows on from previous work which showed that the motion of a spherical object could be controlled by an ordered foam \cite{Davies12}. Again, we simulate the motion of light objects in aqueous foams, so that the effects of gravity do not always dominate the effects of surface tension and a quasi-static regime is appropriate. This is in contrast to the work of Courbin and Stone \cite{Courbin06} and Le Goff \textit{et al.} \cite{LeGoff08} for example, in which soap films and foams were shown to absorb energy from a fast moving object.

Probing a foam's response to solid objects is a standard tool that has been used to develop a better understanding of their complex behaviour. In 2D, Raufaste \textit{et al.} \cite{Raufaste07} performed experiments and simulations and showed that the drag force exerted on a circular object by a flowing foam increases linearly with the diameter of the object, and decreases with increasing foam wetness. Furthermore Dollet \textit{et al.} \cite{Dolletdrag05} showed that the surface properties of the object have a weaker effect on the drag force exerted by the foam. The experiments of Dollet and Graner \cite{DolletGraner07} and our previous 2D bubble-scale simulations \cite{Davies08} found that a flowing foam responds purely elastically far away from a circular obstacle it flows past, and that plastic topological events, known as T1 events, where bubbles swap neighbours, occur mainly within the first couple of layers of bubbles near to the object.

Studying the interaction between 3D foams and solid objects presents challenges in both visualization and computation. Cantat and Pitois \cite{Cantat05,Cantat06}  measure the force exerted on a spherical bead moving at constant velocity through a liquid foam, detecting elastic loading and topological changes, while de Bruyn's experiments \cite{Bruyn04,Bruyn05} confirmed that a fluidized region surrounds a sphere moving through a foam and that the drag force decreases as the average bubble size increases. 

Non-circular or non-spherical objects have also been considered, and the shape of objects (or obstacles) has been shown to greatly affect the response of a foam. In 2D, experiments by Dollet \textit{et al.} \cite{Dolletellipse06} showed that an elliptical object aligns its major axis parallel to the direction of flow of a foam. We also showed, via 2D simulations \cite{Davies10}, that the elasticity of a foam drives an ellipse falling through a foam to rotate to a stable orientation in which its major axis is parallel to gravity. Simulations by Boulogne and Cox \cite{Boulogne11} showed that the drag and lift forces exerted by a flowing 2D foam on different shaped objects is highly dependent on the shape of the object. 

In 3D, the work of Morris \textit{et al.} \cite{Morris10,Morris11,Morris2011} probes the interaction between thin liquid films and super-quadric objects such as cubes or ellipsoids. They investigate the effect that the shape, orientation and surface properties of the solid object have on film stability. Their work, and in particular their simulation methodology is closely related to what we describe in this paper. They prescribe a fixed orientation for the super-quadric object and then use Surface Evolver to find a minimum energy surface for the thin film that surrounds it. Their work investigates the effect that the contact angle of the film has on particle orientation and also what parameter values lead to two sides of films with finite thickness to touch, causing the film to rupture.

In this work, we investigate how the shape and orientation of a particular super-quadric object, a cube with rounded edges and corners, affects the deformation of a soap film that it falls through and the forces that the soap film exerts on the object. We keep the surface properties of the falling objects constant throughout, so that the soap film always meets the object normal to the surface. This can be justified physically by the presence of a wetting film that covers the object. The simulation model and methodology are described in detail in section \ref{sec:method}. The results of our simulations are discussed in section \ref{sec:results}, where we vary the shape of the falling object (\S \ref{sec:shape}), its initial orientation (\S \ref{sec:orientation}) and position relative to the centre of the cylinder containing the film (\S \ref{sec:position}). This is followed by our conclusions and discussion of future work in section \ref{sec:conclusions}.

\section{Method}
	\label{sec:method}

\begin{figure}
  \centering
		\includegraphics[width=0.95\textwidth]{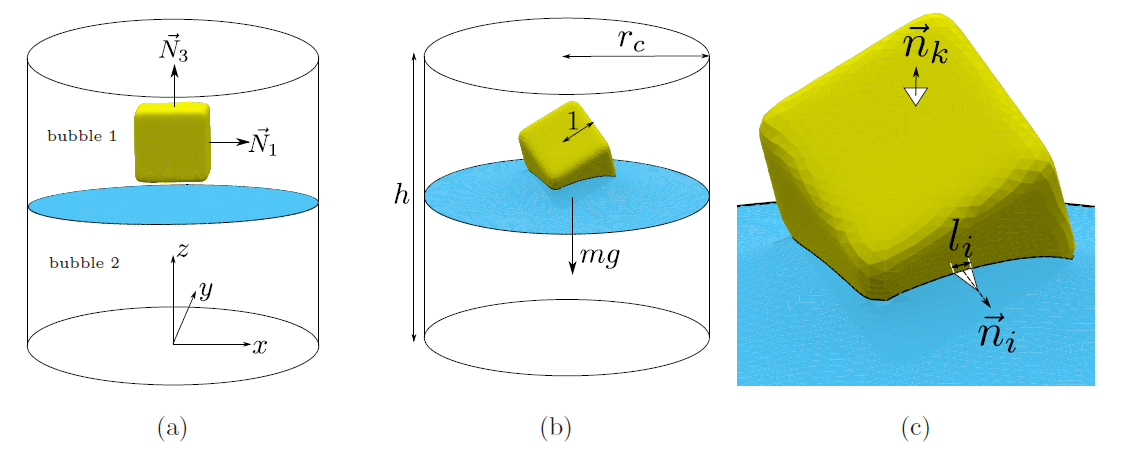}
  \caption{(a) The initial set-up in which a cube is positioned above a horizontal film that separates two bubbles in a cylindrical tube with radius $r_c$ and height $h$. The orientation of the cube is prescribed and then traced by recording the angle that the normal vectors that define it, $\vec{N}_1$, $\vec{N}_2$ (obscured by the cube in the image) and $\vec{N}_3$ make with the $z$-axis. (b) The object (of radius $r_s=1$) is allowed to fall through the soap film under gravity, and therefore deforms the soap film during contact. Its motion is governed by the resultant of its weight, $mg$, and the pressure and network forces that the foam exerts on it. (c) For each edge $i$ of the triangulated surface of the soap film that contacts the object, an outward network force is exerted in the normal direction, $\vec{n}_i$ over its length $l_i$. Similarly, a pressure force is exerted by contacting bubbles, now in the inward normal direction to the surface, $-\vec{n}_k$, over all facets $k$. In the same way, bubble pressures and surface tension contribute towards a network and pressure torque which rotate the object.}
  \label{fig:method}
\end{figure}

The Surface Evolver \cite{Brakke} allows us to resolve bubble pressures and the geometry of thin films for foams under given constraints. Our simulations consists of a single horizontal soap film that separates two bubbles contained in a cylindrical tube of radius $r_c$, height $h$ and a vertical centre-line that coincides with the $z$-axis of the Cartesian coordinates, so that the centre of the cylinder's base lies at the origin. 

The surface of the solid object that falls through the soap film is defined by the super-quadric equation
\begin{equation}
\left(x-x_0\right)^{\lambda}+\left(y-y_0\right)^{\lambda}+\left(z-z_0\right)^{\lambda}=r_s^{\lambda},
\label{eq:superquadric}
\end{equation}
where $\left(x_0,y_0,z_0\right)$ denotes its centre coordinates, $r_s$ its radius (which we set to be equal to one throughout this work), and $\lambda$ is a shape parameter that satisfies $\lambda=2n$ where $n\in\mathbb{N}^+$. Note that when $\lambda=2$ the object is a sphere with radius $r_s=1$, while increasing $\lambda$ yields a cube with rounded edges and corners \cite{Jaklic00,Zhou01}. For a cube, $r_s=1$ describes the minimum distance between its centre coordinates and its surface, and for convenience we shall refer to this as the radius of the cube for the rest of this paper. This solid object is initially positioned well above the soap film so that they do not intersect (see figure \ref{fig:method}a).

The surface tension of the soap film is denoted by $2\gamma$, and this is set to be equal to $1$ throughout this work. The volume of the $k-th$ bubble (where $k=1,2$) is denoted by $V_k$ while $V_k^t$ is a target volume assigned to each bubble. The soap film, a triangulated mesh, is equilibrated by minimizing its surface area $A$, using the energy functional, $E$:
\begin{equation}
E=2\gamma A + \sum_{k} p_k\left(V_k-V_k^t\right),
\label{eq:energy}
\end{equation}
where $p_k$ is a Lagrange multiplier that denotes the pressure of the bubbles. Note that both bubbles are assigned the same target volume
\begin{equation}
V_k^t=\frac{1}{2}\left(\mbox{volume of cylinder - volume of solid object}\right)=\frac{1}{2}\left(2\pi r_c h-\frac{8}{3\lambda^2}\frac{\Gamma\left(\frac{1}{\lambda}\right)^3}{\Gamma\left(\frac{3}{\lambda}\right)}\right).
\end{equation}
The calculation of the volume of the super-quadric object involves the numerical computation of gamma functions, for example $\Gamma\left(1/\lambda\right)=\int_0^\infty t^{1/\lambda-1}e^{-t}dt$.   

We set the initial orientation of the object by rotating it by prescribed angles $\phi_j$, $\theta_j$ and $\psi_j$ around the $x$, $y$ and $z$ axes respectively.    This involves multiplying the surface constraint given in equation \ref{eq:superquadric} by the rotation matrices
\begin{equation}
\textbf{R}_{x,j} = \left( \begin{array}{ccc}
1 & 0 & 0 \\
0 & \cos\phi_j & -\sin\phi_j \\
0 & \sin\phi_j & \cos\phi_j \end{array} \right),\,\,
\textbf{R}_{y,j} = \left( \begin{array}{ccc}
\cos\theta_j & 0 & \sin\theta_j \\
0 & 1 & 0 \\
-\sin\theta_j & 0 & \cos\theta_j \end{array} \right),\,\,
\textbf{R}_{z,j} = \left( \begin{array}{ccc}
\cos\psi_j & -\sin\psi_j & 0 \\
\sin\psi_j & \cos\psi_j & 0 \\
0 & 0 & 1 \end{array} \right),\,\,
\label{eq:rotation_matrix}
\end{equation}
in this order. Note that this is equivalent to applying the combined rotation matrix
\begin{equation}
\textbf{R}_j = \textbf{R}_{z,j}\textbf{R}_{y,j}\textbf{R}_{x,j}.
\label{eq:rotation_matrix1} 
\end{equation}
Note that we include an index $j$, which is equal to zero during this initial setup, as we will also apply this rotation matrix at each iteration of our simulation (see later).

The object is assigned a weight $mg$ and allowed to descend through the soap film under gravity. As in previous work \cite{Davies12}, we use a quasi-static model, which assumes that the motion of the object through the soap film is slow. A further simplification is made by assuming that this motion is also steady. This model is only appropriate when the object is in contact with the soap film, and this is therefore the focus of this work. The position of the object is lowered until its surface becomes within a critical separation of $0.05$ of the soap film. At this stage, the facets of the soap film and the object that are closest together are merged and the soap film is equilibrated using a combination of gradient descent and conjugate gradient energy minimization steps. These are inter-dispersed with tests for detachment from the object and upkeep of the tessellation. The minimisation procedure continues until convergence to within a tolerance of $1\times 10^{-5}$ has been achieved. The super-quadric object and the cylindrical container are assumed to be covered by a wetting film and the soap film is free to slip along both surfaces, and as a consequence of energy minimization, will always meet them at $90^\circ$ (see figure \ref{fig:method}). 

The contacting soap film exerts a network force, $\vec{F}^n$ on the solid object due to the pull of surface tension. This force is calculated geometrically as
\begin{equation}
\vec{F}^n=2\gamma\sum_i l_i\vec{n}_i
\label{eq:networforce}
\end{equation}
where $l_i$ denotes the contact length of the triangular facet $i$ of the soap film that is in contact with the object and $\vec{n}_i$ denotes the unit normal vector to the surface at $\left(x_i,y_i,z_i\right)$, the mid-point of edge $i$ (see figure \ref{fig:method}c). Since the surface contacts the object at $90^\circ$, the pull it exerts on it contributes towards a network torque, $\vec{\tau}^n$. Letting $\vec{r}_i=\left(x_i-x_0,y_i-y_0,z_i-z_0\right)$ denote the vector that connects the centre of the object with the midpoint of edge $i$, it is clear that $\vec{r}_i$ and $\vec{n}_i$ are in general not parallel. The network torque is calculated as the sum of vector cross-products 
\begin{equation}
\vec{\tau}^n=2\gamma\sum_i l_i\vec{r}_i\times\vec{n}_i.
\label{eq:networktorque}
\end{equation}
Similarly, bubbles in contact with the object apply a pressure force, $\vec{F}^p$, over its surface and this is calculated by the summation 
\begin{equation}
\vec{F}^p=-\sum_k p_k A_k\vec{n}_k
\label{eq:pressureforce}
\end{equation}
where  $A_k$ denotes the area of the $k$th facet of the object, and $\vec{n}_k$ denotes the outward unit normal vector positioned at the midpoint of this facet, say $\left(x_k,y_k,z_k\right)$. Note that the negative sign is due to the fact that the bubble applies an inward push due to its pressure, $p_k$. As before, there is a contribution from this force towards a pressure torque, $\vec{\tau}^p$. Let $\vec{r}_k=\left(x_k-x_0,y_k-y_0,z_k-z_0\right)$ denote the vector that connects the centre coordinates of the object with the mid-point of the $k$-th triangular facet of the object. As before $\vec{r}_k$ and $\vec{n}_k$ are not necessarily parallel, and therefore the calculation for pressure torque is given by the summation
\begin{equation}
\vec{\tau}^p=-\sum_k A_kp_k\vec{r}_k\times\vec{n}_k.
\label{eq:pressuretorque}
\end{equation}
Therefore the resultant force and torque exerted on the super-quadric object are
\begin{eqnarray}
\vec{F}&=&-mg\vec{z}+\vec{F}^n+\vec{F}^p,\\
\vec{\tau}&=&\vec{\tau}^n+\vec{\tau}^p,
\end{eqnarray}
respectively, where $\vec{z}$ denotes an unit vector in the positive $z$ direction. We will from now on use the component form of these forces, that is $\vec{F}=\left(F_x,F_y,F_z\right)$ and $\vec{\tau}=\left(\tau_x,\tau_y,\tau_z\right)$.

Each simulation iteration involves equilibrating the soap film while the object's position and orientation is fixed, calculating the forces it exerts on the object, and then moving the object in the direction of the resultant force by a small amount. We choose a small constant $\varepsilon$ that sets the effective time-scale of our simulations. At each iteration we move the object by $\varepsilon\vec{F}$ and rotate it by $\varepsilon\vec{\tau}$, using the standard right hand convention for rotation. This requires applying a rotation matrix such as that given in equation \ref{eq:rotation_matrix1} where the angles of rotation around the $x$, $y$ and $z$ axes are $\phi_j=\varepsilon\vec{\tau}_x$, $\theta_j=\varepsilon\vec{\tau}_y$ and $\psi_j=\varepsilon\vec{\tau}_z$ respectively for the $j$th iteration, where $j=1,2,3,...\,$. Thus after $n$ iterations, the orientation of the object is given by the $3\times 3$ matrix 
\begin{equation}
\textbf{R} = \prod_{j=n}^{j=0} \textbf{R}_j= \left( \begin{array}{ccc}
r_{1,1} & r_{1,2} & r_{1,3} \\
r_{2,1} & r_{2,2} & r_{2,3} \\
r_{3,1} & r_{3,2} & r_{3,3} \end{array} \right).
\label{eq:orientation}
\end{equation}
The columns of this matrix can be thought of as the unit normal vectors that define the orientation of the super-quadric object, which we denote by $\vec{N}_1$, $\vec{N}_2$ and ${N}_3$ respectively (see figure \ref{fig:method}a). We will record the orientation of the falling object by determining the angles that these three vectors make with the vertical $z$-axis, which we denote by $\alpha_1=\cos^{-1}\left(r_{3,1}\right)$, $\alpha_2=\cos^{-1}\left(r_{3,2}\right)$ and $\alpha_3=\cos^{-1}\left(r_{3,3}\right)$ respectively.  

Throughout this work, we fix the density, $\rho$, of the object so that the Bond number, $Bo=\rho g r_s^2/\gamma=1.95$ (to two decimal places). This density has been chosen to ensure that the object detaches itself from the soap film and cannot be supported by the soap film in any orientation. We also set $\varepsilon=1/400Bo$, which ensures convergence in the sense that the results do not change by making $\varepsilon$ smaller. The equilibration of the soap film involves performing up to 10,000 gradient descent and conjugate gradient minimization steps interspersed with upkeep of the tessellation and checks for soap film detachment. We also apply small perturbations to the surface of the soap film during the equilibration process by jiggling the vertices slightly. This ensures that the surface is converging towards an absolute minimum energy, and not a local minimum. The equilibration process is terminated once the surface area of the soap film has converged to five decimal places.

\section{Results}
 \label{sec:results}

Although the simulation that we model is relatively simple, it provides a rich system to study as we can vary many parameters and investigate the effect of doing so on the interaction that occurs between a falling solid object and a soap film. In this section, we summarise the results of varying the shape, initial orientation and position of the object. 

\subsection{Variation of shape}
	\label{sec:shape}

The first parameter that we vary is $\lambda$, the shape parameter from equation \ref{eq:superquadric}. We increase $\lambda$ in steps of two between $\lambda=2$ (a sphere) and $\lambda=20$ (a cube with smooth rounded edges). Throughout this section the radius of the object is kept fixed at $r_s=1$ while the cylinder has radius $r_c=4$. The initial orientation of the object is also kept constant in this section, with the angles used for setting up the orientation of the object all set to zero so that the cube presents a flat face to the soap film. The object is free to rotate as it falls through the soap film. The initial position of the object is also kept constant so that the centre coordinates of the object lie at $x_0=y_0=0$ and $z_0>6$, that is at the centre of the cylinder and above the soap film. 

\begin{figure}
  \centering
		\includegraphics[width=0.85\textwidth]{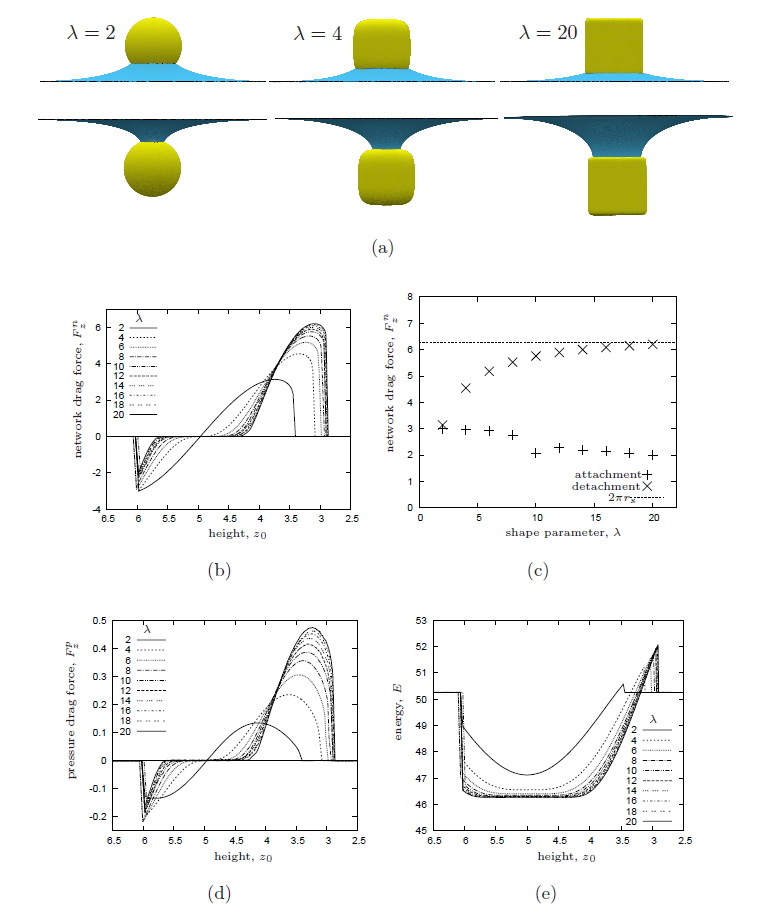}
  \caption{Variation of the shape parameter $\lambda$: (a) A side view of the soap film's deformation after attachment (top) and before detachment (bottom) from objects with $\lambda=2,\,4\,\mbox{and}\,20$ (from left to right). (b) The network drag force, $F_z^n$,  exerted on the objects as they fall through a soap film versus their height, $z_0$. (c) The absolute value of the network drag force straight after soap film attachment and just before detachment. (d) The pressure drag force, $F_z^p$,  exerted on the objects as they fall through a soap film versus their height, $z_0$. (e) The energy $E$ (i.e. surface area) of the soap film as it is penetrated by the objects versus the height of the object in the cylindrical container.}
  \label{fig:shape_variation}
\end{figure}

We focus on the deformation caused to the soap film by the falling object, and the forces exerted by the soap film on the object as a consequence. Snapshots of the simulation just after the film attaches itself to the falling object and just before it detaches are given in figure \ref{fig:shape_variation}a. These show how the deformation differs for three examples; a sphere ($\lambda=2$), an object that is between a sphere and a cube ($\lambda=4$) and a cube with smooth rounded edges ($\lambda=20$). 

The shape of the falling object clearly affects how the soap film deforms. After attachment, the contact line between the soap film and the object evolves to its highest position in the case of a sphere. This is a direct result of energy minimization, as the soap film contacts the object at 90$^{\circ}$ to the surface at its minimum energy. Just after attachment to the sphere, the soap film bends upwards to find its minimal surface area while still satisfying the bubble volume constraints. However, when the soap film contacts the cubic objects (where $\lambda=4$ and $\lambda=20$), the contact line of the soap film does not have to rise far before finding its minimum energy. The surface normals of the object around its four vertical faces are closer to being horizontal as $\lambda$ increases,  meaning that they are close to being parallel to the orientation of the soap film prior to attachment. As a result, the positional evolution of the soap film's surface during energy minimization after attachment is smaller. It is noticeable from our simulations that the deformation of the soap film during attachment decreases for increasing $\lambda$. 

Conversely, it is clear from figure \ref{fig:shape_variation}a that the soap film's deformation prior to detachment from the falling object increases with $\lambda$. The rounded shape of a sphere means that the contact line of the soap film moves steadily up over its surface as it descends, resulting in the smoothest of detachments. However, more cubic objects have flatter surfaces separated by regions of high curvature. Thus, just before detachment, the soap film's contact angle with the object is such that it is near vertical. Once the soap film has reached this vertical configuration (see figure \ref{fig:shape_variation}a, bottom right), and the solid object is allowed to fall further, the energy minimization of the soap film results in detachment from the object. The sharper corner of a cubic object means that it takes longer for the soap film to reach the point at which detachment occurs, and as a consequence it becomes more stretched than for a sphere. Of course, when the angles that set the orientation of the cube are non-zero, the soap film deforms less while navigating the uppermost slanted surfaces of the cube before detaching (see later).

Figure \ref{fig:shape_variation}b shows how the network drag, $\vec{F}^n_z$, varies with the height of the super-quadric object in the cylinder for different values of $\lambda$. Note that the height of the soap film in the cylinder before it attaches itself to the falling object is close to $z=5$. Once the object's centre reaches a height of $z_0\approx 6$, its lower surface is within touching distance of the soap film, and therefore attachment occurs close to this point. The attachment of the soap film to the object is shown in figure \ref{fig:shape_variation}b by a sharp negative (downward) network drag, which is a result of the angle at which the soap film contacts the object after attachment. The magnitude of the network drag after the soap film attaches itself to the solid object seems to be independent of $\lambda$. This is surprising considering that the deformation after attachment clearly varies with $\lambda$. The lack of variation of the network drag after attachment for different values of $\lambda$ may be explained by the fact that the contact angle the soap film makes with the horizontal plane is balanced out by its contact length over the object. For example, when the soap film attaches itself to the sphere, it becomes highly deformed, thus contacting the sphere at a large angle to the horizontal, but the length of this contact line is small. Conversely the contact line of the soap film with a cube is clearly longer after attachment, but the soap film is less deformed, meaning the angle it makes with the horizontal is small. The non-variation of this network drag after attachment with $\lambda$ is confirmed in figure \ref{fig:shape_variation}c. Once the soap film is attached to the object, the network force increases to zero when the position of the object's centre in the cylinder, $z_0$, is equal to the height at which the outer edges of the soap film touch the cylinder wall. Once the object passes this point, the network force becomes positive, thus contributing to resisting the downward motion of the object. The maximum network drag is achieved for the most cubic object (where $\lambda=20$). Note that as well as achieving a maximum drag, the interval of height for which the object is in contact with the super-quadric object increases with $\lambda$, as the position of the soap film's contact line with the object stagnates on the rounded edges of the cube's upper face. As these rounded edges become more curved as $\lambda$ increases, the minimal surface remains in contact with the edges for more iterations, and the soap film becomes more stretched in the vertical direction before detaching from the object. This result is confirmed in figure \ref{fig:shape_variation}c, which shows that the absolute value of the network drag instantaneously before the soap film detaches from the object increases with $\lambda$. We propose that under the given conditions, the network drag tends to $2\pi r_s$ as $\lambda$ increases without bound (for a surface tension of $2\gamma=1$). This is the circumference of the largest circle that can be inscribed within the upper square surface of the cube. Once the film is no longer contacting the rounded edges of the cube, its surfaces slips along the flat surface of the cube during the minimization process before detaching.      

The pressure drag, $\vec{F}^p_z$, versus the height, $z_0$, of the object in the cylinder is shown in figure \ref{fig:shape_variation}d for the same set of values of $\lambda$. It is noticeable here that the pressure contribution to the drag force is an order of magnitude smaller than that of the network force, and that it acts in the same direction. This was also seen to be the case in the simulations of Davies and Cox \cite{Davies12}. The pressure drag exerted on a sphere is smaller in magnitude, undergoes a smoother transition and exhibits more symmetry between attachment and detachment compared with the pressure drag for objects where $\lambda>2$. Recall that the pressure force relates to the pressure difference between the two bubbles, and therefore the curvature of the film that separates them. After attachment, the curvature of the soap film is such that the pressure in the lower bubble is less than the pressure of the upper bubble. In this case, the pressure drag force is negative during the attachment (and therefore it contributes to drag the object downwards). As for the network force, it increases to zero when the centre of the object is perfectly aligned with the position of the soap film. This is to be expected as the shape of the soap film is such that its curvature is zero here, and therefore the pressure difference between the two bubbles is zero. As the object descends further, the curvature of the soap film switches sign so that the pressure drag becomes positive, increasing until reaching a maximum value just before detachment. The maximum pressure drag exerted increases with $\lambda$ in a similar fashion to the network contribution to the drag force.

Figure \ref{fig:shape_variation}e shows how the surface area of the soap film varies with the height of the centre of the object in the cylinder for different values of $\lambda$. Before the soap film attaches to the object, its area is simply $2\pi r_c$. That area sharply decreases after the soap film attaches itself to the object, reaching a minimum when the height of the centre of the object in the cylinder is perfectly aligned with the position of the soap film. The cross-sectional area of the object is at its largest here for all values of $\lambda$, and therefore it is to be expected that the minimum energy is attained here. The soap film is then stretched by the object as it falls beyond this point, with a maximum energy reached just before detachment. Again, is is clear from this figure that the amount of film stretching required before detachment increases with $\lambda$.

\subsection{Variation of initial orientation}
	\label{sec:orientation}

In this section we inspect how the orientation of the object affects the deformation it causes to the soap film it falls through. To isolate the effect of the object's orientation, we keep the shape parameter of the object fixed at $\lambda=10$. We vary the initial orientation of the cube by choosing different values for $\phi_0$ and $\theta_0$, the angles by which the object is rotated around the $x$ and $y$ axes during the initial setup respectively. Note that we keep $\psi_0$, the angle of rotation around the $z$ axis, equal to zero. Note that we're allowing the cube to fall from the centre of the cylinder, so varying $\psi_0$ will not affect the result due to the symmetry of the problem around the centreline of the cylinder.  We inspect how the soap film affects the orientation of the object, investigating whether or not a bamboo foam can be used to re-orient a cubic object in a controlled and predictable way. Recall that we record the orientation of the object by $\alpha_1$, $\alpha_2$ and $\alpha_3$, the angles that the normal vectors that define the orientation of the object, $\vec{N}_1$, $\vec{N}_2$ and $\vec{N}_3$, make with the $z$-axis (see figure \ref{fig:method}a).

\begin{figure}
    \centering
			\includegraphics[width=0.95\textwidth]{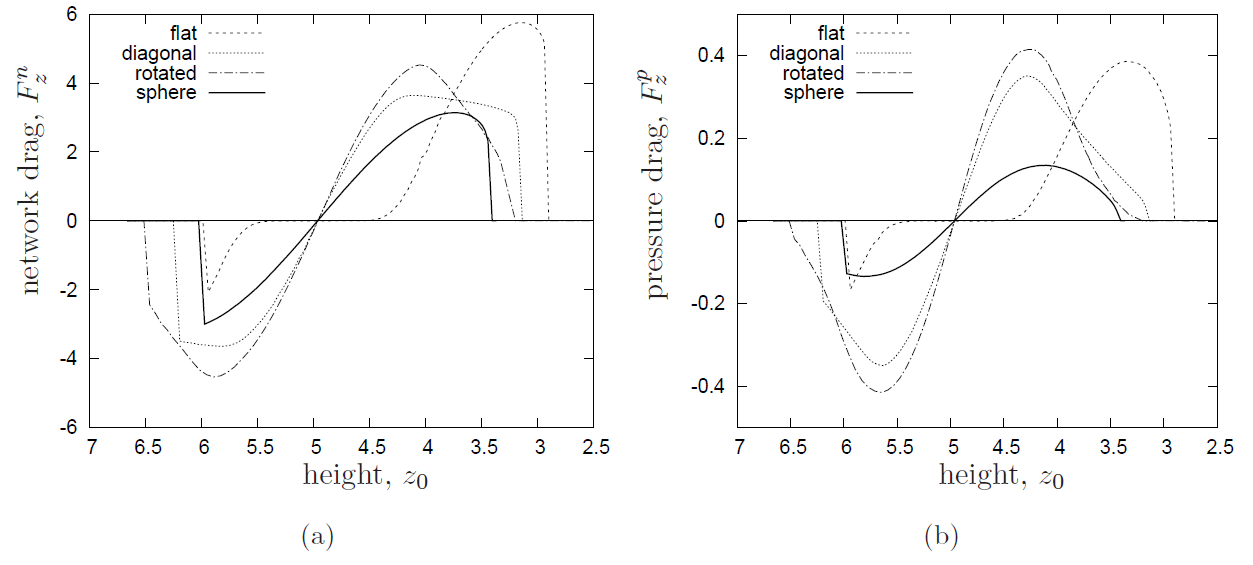}
		\caption{The drag force exerted on a cube falling through the centre of a soap film in three orientations: \emph{Flat} where the angles between the normal vectors defining the cube, $\vec{N}_1$, $\vec{N}_2$ and $\vec{N}_3$ and the $z$ axis are initially $\alpha_1=\alpha_2=\pi/2$ and $\alpha_3=0$ respectively, \emph{diagonal} where $\alpha_1=\alpha_2=\pi/4$ and $\alpha_3=\pi/2$, and \emph{rotated} where $\alpha_1=\alpha_2=\alpha_3=\pi/2-\tan^{-1}\left(1/\sqrt{2}\right)$. The drag force is separated into (a) a network component, $F^n_z$, and (b) a pressure component, $F^p_z$, and plotted versus the height, $z_0$, of the object in the cylindrical container. Note that the forces exerted on a sphere are also included for comparison.}
	\label{fig:drag_stable}
	\end{figure}

\subsubsection{Stable orientations}

\begin{figure}
	\centering
\includegraphics[width=0.95\textwidth]{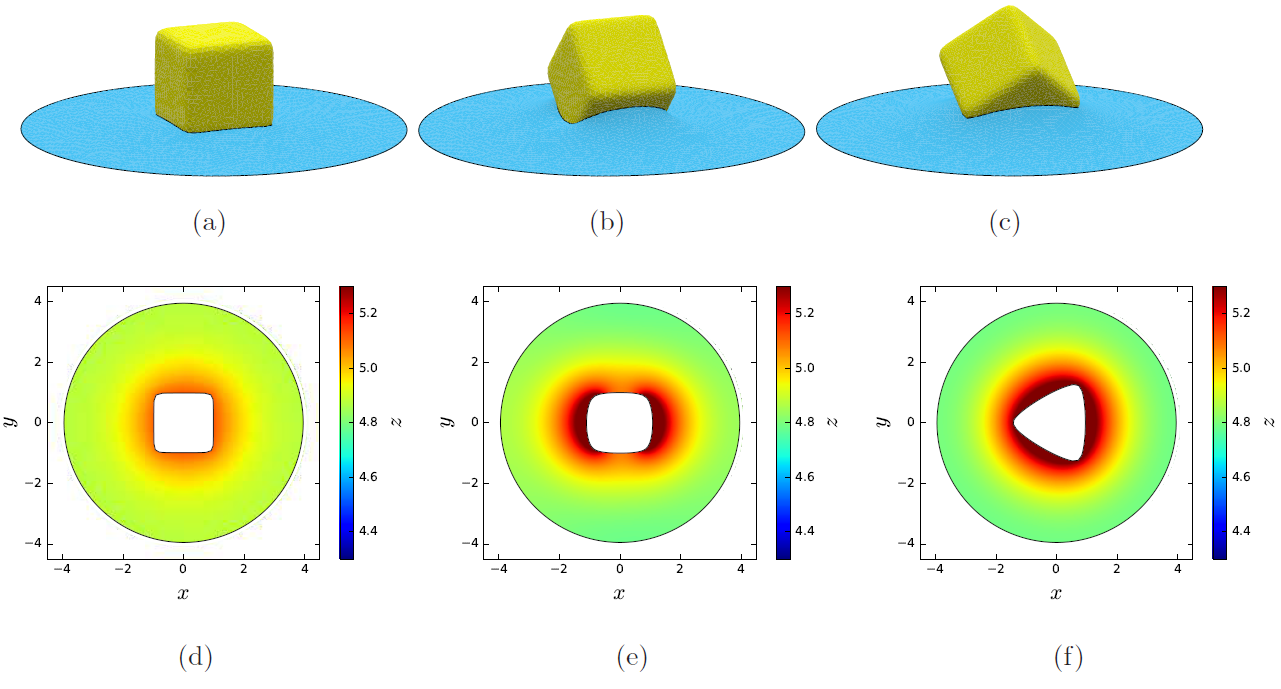}
	\caption{Snapshots of the simulation when a (a) \emph{flat}, (b) \emph{diagonal} and (c) a \emph{rotated} cube (with $\lambda=10$) reaches a height of $z_0\approx 5.9$ in the centre of the cylinder, thus contacting and deforming the soap film. The shape of the surface of the film for these three cases is visualized by surface plots directly below the snapshots in (d), (e) and (f) respectively, with the height of the film indicated by colour.}
	\label{fig:cube_attachment}
\end{figure}

Let us first consider three initial orientations of the cube which do not change as the cube falls through the centre of the soap film. In fact we have already considered the first of these orientations in the previous section. In that case the angles between the normal vectors $\vec{N}_1$, $\vec{N}_2$ and $\vec{N}_3$ and the $z$ axis are $\alpha_1=\alpha_2=\pi/2$ and $\alpha_3=0$ respectively. We will denote this as the \emph{flat} orientation from now on. The next orientation that we consider is the \emph{diagonal} one, where $\alpha_1=\alpha_2=\pi/4$ and $\alpha_3=\pi/2$. The third orientation is set so that all three normal vectors defining the cube have the same angle with the $z$ axis, that is $\alpha_1=\alpha_2=\alpha_3=\pi/2-\tan^{-1}\left(1/\sqrt{2}\right)$, and we call this the \emph{rotated} orientation. These three orientations are shown in figure \ref{fig:cube_attachment} just after the soap film has attached to the cube. Their stability when the cube falls through the centre of the foam can be explained by the symmetry of the problem. For these three orientations, the deformation of the soap film will be symmetric around the $z$ axis, meaning the torque exerted by the foam on the cube will be negligible. In fact, we will show that the \emph{flat} orientation is the only stable orientation for a cube falling down the centre of a bamboo foam, and that the \emph{diagonal} and \emph{rotated} orientations are meta-stable.

\begin{figure}
	\centering
\includegraphics[width=0.95\textwidth]{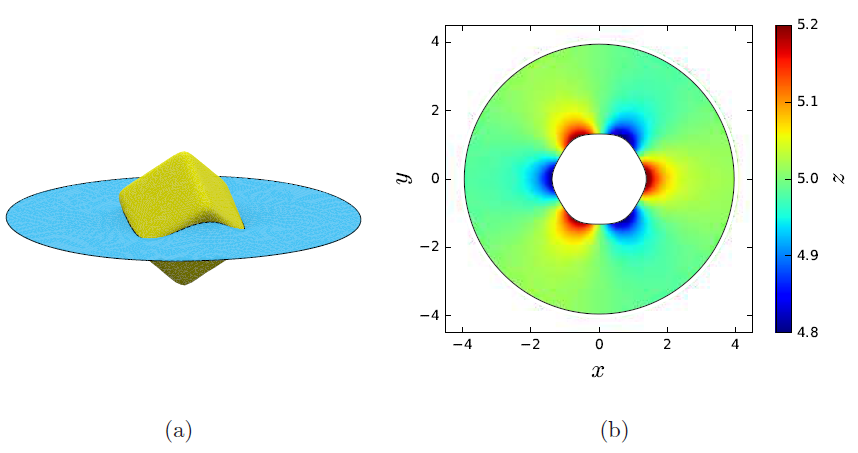}
	\caption{soap film deformation when the centre coordinates of a cube in the \emph{rotated} orientation are aligned with the position of the soap film: (a) Simulation snapshot and (b) surface plot of the height of the soap film.}
	\label{fig:cube_immersed}
\end{figure}

\begin{figure}
\centering
\includegraphics[width=0.95\textwidth]{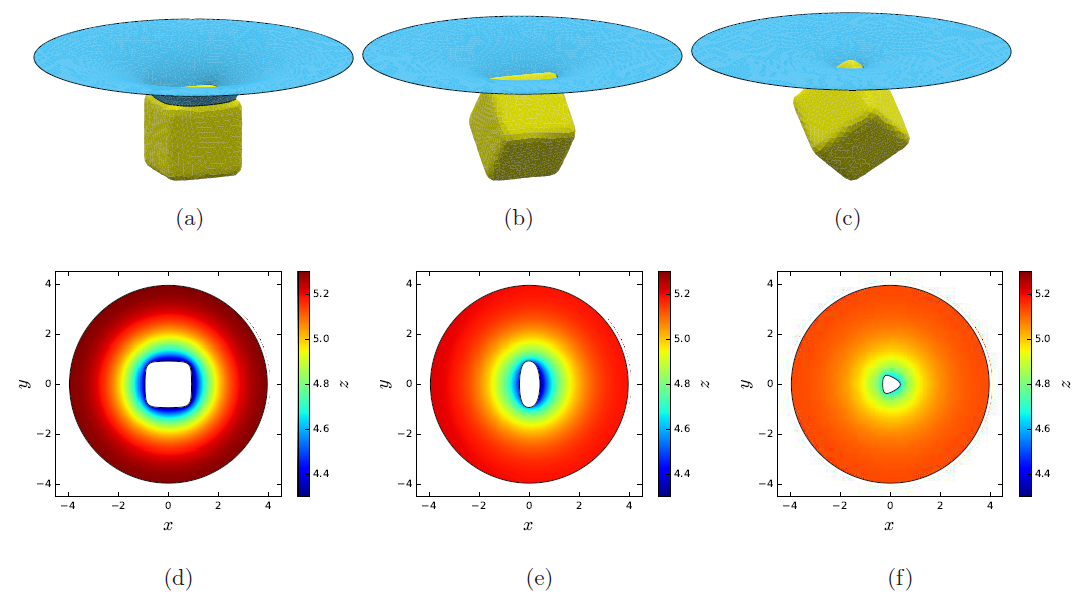}
	\caption{Snapshots of the simulation when the centre of a cube reaches a height of $z_0\approx 3.4$ in the centre of the cylinder, thus getting close to detaching from the film in the (a) \emph{flat}, (b) \emph{diagonal} and (c) \emph{rotated} orientations. The shape of the soap film for these three snapshots is visualized by surface plots in (d), (e) and (f) respectively, with the height of the surface denoted by colour.}
	\label{fig:detachment_comparison}
\end{figure}

Figure \ref{fig:drag_stable} shows the drag force the soap film exerts on the cube in these three orientations. The network component of the drag force is shown in figure \ref{fig:drag_stable}a, where it is clear how important the object's orientation is. Once the soap film attaches itself to the cube, the network force can be seen to jump from zero to a negative value, showing a downward pull. This jump is shortest for the rotated cube, where the soap film finds its minimal area without deforming as much, only needing to engulf the leading apex of the cube.  The initial network force exerted on the diagonal cube is higher than for any other orientation, as in this case the soap film contacts the object over a greater length than for the rotated cube and at a larger angle to the horizontal than for the flat cube (see figure \ref{fig:cube_attachment}).

As the cube falls further, the downward network force decreases to a minimum value for the rotated cube, where the cross-sectional area to be navigated by the soap film is largest. The negative network drag is also greater in magnitude for the diagonal cube compared to the flat cube, as the angle the soap film makes with the horizontal here while contacting the cube is larger. The network drag then increases to zero as $z_0$ approaches the vertical position of the soap film. Here, the soap film becomes completely horizontal for the diagonal and flat cube. However, for the rotated cube it takes a hexapolar deformation that includes three rises and three depressions that are symmetric around the $z$-axis (see figure \ref{fig:cube_immersed}). This deformation was seen for a cube in thin films in the work of Morris \textit{et al.} \cite{Morris11} and for cubes lying at fluid-fluid interfaces in the work of Soligno \textit{et al.} \cite{Soligno16}.

Figure \ref{fig:drag_stable}a shows that out of the three orientations, it is the rotated cube that experiences the largest network drag force, which is to be expected as it has the largest cross sectional area meaning that the soap film will contact it over a greater length. In this case, the network drag diminishes steadily after reaching a maximum, due to the fact that once the soap film has passed the point shown in figure \ref{fig:cube_immersed}, it is free to slip along the slanted surfaces of the cube and detach. The network drag force exerted on the diagonal cube at the same stage is smaller in magnitude, but is applied on the cube over a prolonged interval of height. Again, this can be explained by the geometry of the problem; the soap film has to navigate two corners of the cube where the curvature is high. Navigating corners or edges slows down the process of film detachment in a similar but less extreme way than what was seen for the flat cube. The deformation caused to the soap film just before it detaches from the cube is shown in figure \ref{fig:detachment_comparison}. Note that the vertical position of the film is higher in figure \ref{fig:detachment_comparison} compared to its position in figure \ref{fig:cube_attachment}. This is a result of the cube penetrating the lower bubble of the two, which forces the film to move upwards so that the volume constraints imposed are still satisfied.

More insight can be gained by looking at how the pressure drag force varies with the height of the cube in the cylinder (see figure \ref{fig:drag_stable}b). The most striking result here is how antisymmetric the pressure drag force is between when the soap film attaches to and detaches from the cube for the rotated and diagonal orientations. We previously discussed how the soap film deformation was not symmetric at all for the flat cube. We note that there is some antisymmetry for the case of a sphere, but it is closer to full antisymmetry for the rotated and diagonal cubes. For these orientations, the deformation caused to the soap film by the cube is symmetric about the point at which the cube's centre coordinates are perfectly aligned with the position of the soap film. Note that the direction of the curvature of the film's surface before detachment is the opposite of the direction of curvature after attachment, but that the overall geometry is very similar for these orientations (see figures \ref{fig:cube_attachment}e,f and \ref{fig:detachment_comparison}e,f).

\begin{figure}
    \centering
\includegraphics[width=0.95\textwidth]{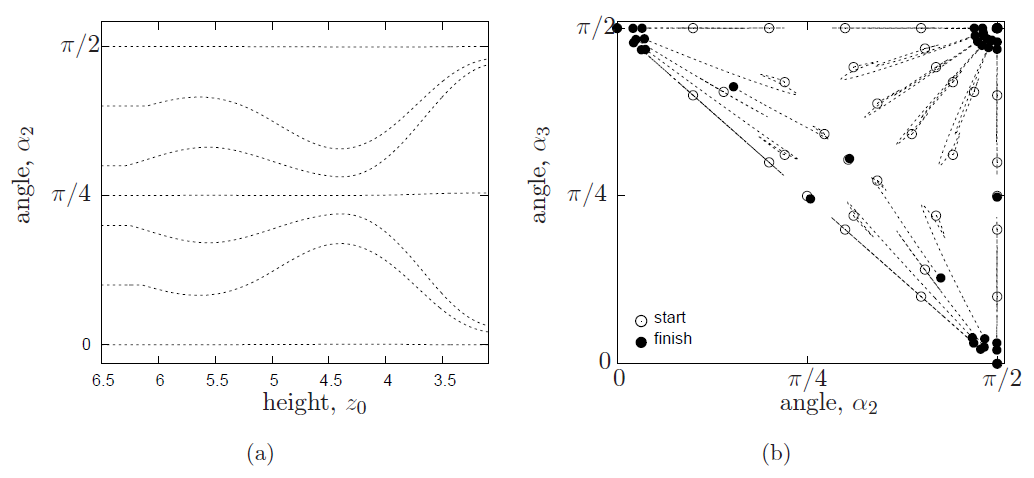}
		\caption{(a) Variation of the orientation of the cube for different initial values of $\alpha_2$ versus the height of the cube in the cylindrical tube. Here $\alpha_1$ is set to $\pi/2$ initially and therefore it follows that $\alpha_3=\pi/2-\alpha_2$. (b) Variation of the orientation of the cube for different initial orientations. Here, we vary the initial orientation of the cube by considering every possible pairwise combination of $\phi_0$ and $\theta_0$ from the set of values $\left\{0,0.1\pi,0.2\pi,0.3\pi,0.4\pi,0.5\pi\right\}$. The orientation of the cube is plotted in terms of the angle $\alpha_2$ versus $\alpha_3$. An empty circle represents the starting orientation of the cube, a dashed line shows how the orientation varies throughout the simulation and the filled circle shows the final orientation of the cube after it has detached from the soap film.}
			\label{fig:orientation}
	\end{figure}

\subsubsection{Unstable orientations}

Let us now consider initial orientations for the cube which cause the soap film to deform in a non-symmetric way, and therefore where the network and pressure torques exerted by the foam become non-negligible. We first consider rotations of the cube around only one axis, by fixing $\phi_0=\psi_0=0$ and varying the value of $\theta_0$ during the initial setup between zero and $\pi/2$. Recall that we record the orientation of the cube using the angles $\alpha_1$, $\alpha_2$ and $\alpha_3$, that the three normal vectors $\vec{N}_1$, $\vec{N}_2$ and $\vec{N}_3$ make with the $z$-axis  respectively. In this case, the angle $\alpha_1$ is initially equal to $\pi/2$ while $\alpha_2$ is varied and $\alpha_3=\pi/2-\alpha_2$. Figure \ref{fig:orientation}a shows how $\alpha_2$ varies for the cube from different initial values versus its height in the foam. Note that when $\alpha_2$ is initially zero or $\pi/2$, the cube is in the \emph{flat} stable orientation, which we investigated in the previous section. The stability of this orientation is confirmed by the fact that $\alpha_2$ does not change with height for these two simulations. Similarly, when the initial value of $\alpha_2$ is $\pi/4$, we have the \emph{diagonal} orientation, and again the angle recorded here does not change as the cube interacts with the soap film. 

\begin{figure}
\centering
\includegraphics[width=0.8\textwidth]{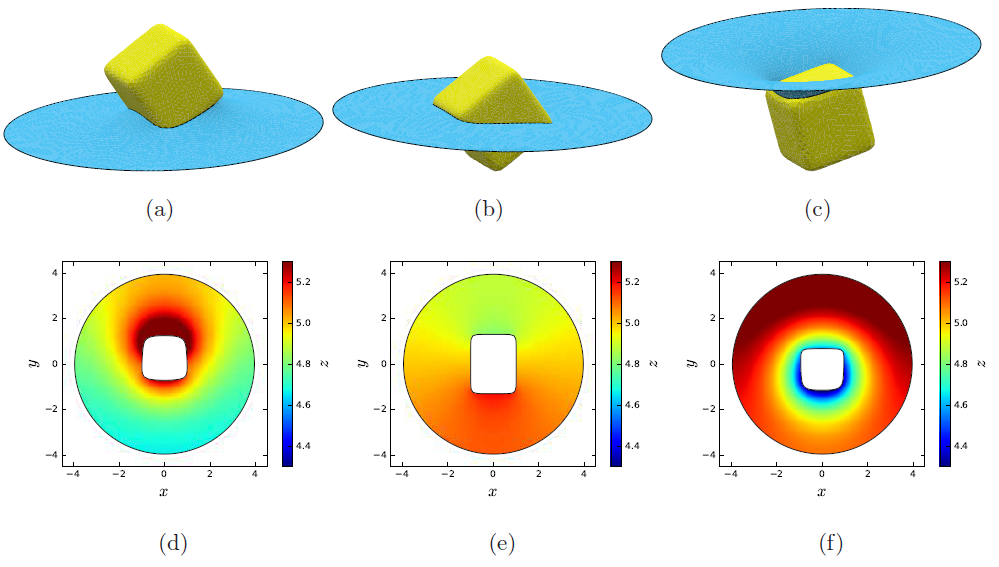}
	\caption{Snapshots of the simulation when the centre of a cube that is initially oriented so that $\alpha_1=0.5\pi$, $\alpha_2=0.3\pi$ and $\alpha_3=0.2\pi$, reaches a height of (a) $z_0\approx 6$, (b) $z_0\approx 5$ and (c) $z_0\approx 3.5$. The shape of the soap film for these three snapshots is visualized by the surface plots in (d), (e) and (f) respectively, with the height of the surface denoted by colour.}
	\label{fig:off_diagonal_sim}
\end{figure}

The new and interesting result here is what happens in between these two orientations. As expected from symmetry, the results for $\alpha_2=0.1\pi$ and $\alpha_2=0.4\pi$ are equivalent, as are the results for $\alpha_2=0.2\pi$ and $\alpha_2=0.3\pi$. In all of these cases, when the soap film attaches itself to the cube it exerts a slight non-zero torque that acts to rotate the object towards the \emph{flat} stable orientation. Here, the soap film moves above the lowest side of the cube's lower face, but remains below the opposite edge of the same face (see figure \ref{fig:off_diagonal_sim}a). As a result the shape of the soap film is not symmetric around the $z$-axis, as shown in figure \ref{fig:off_diagonal_sim}d. In figure \ref{fig:off_diagonal_sim}a, the contribution to the torque exerted by the film on the cube is largest on the right hand side, where the film reaches its highest point. This contributes to rotate the cube in figure \ref{fig:off_diagonal_sim}a in the clockwise direction, and therefore towards the \emph{flat} orientation. 
This is cancelled out by an opposite torque that is applied after the cube falls further and its centre coordinates become more aligned with the vertical position of the soap film. It is shown in figure \ref{fig:off_diagonal_sim}b and \ref{fig:off_diagonal_sim}e that the film is slightly higher on the opposite side to what it was previously. However, as the soap film slips further along the surface of the cube the torque that the foam exerts increases dramatically. The soap film slips towards the rounded edge of the cube that surrounds its upper face (see figures \ref{fig:off_diagonal_sim}c and f). This rounded edge is hard for the soap film to navigate and it is therefore in contact with it over more iterations, exerting a network torque that in effect rotates the cube strongly towards the flat orientation. For example, in figure \ref{fig:off_diagonal_sim}c, the contribution to the torque is much higher from the left hand side as the film contacts the cube nearly vertically here, thus applying a large upward pull on this side. It is clear from figure \ref{fig:orientation}a that the \emph{flat} stable orientation is the most likely outcome of dropping a cube that has been rotated around only one axis through a soap film, and that it is the highly non-symmetric deformation of the soap film that occurs before detachment that is the main driving force for this result.

We now go on to vary the values of both the angles $\phi_0$ and $\theta_0$ used to set the initial orientation of the cube, doing so in increments of $\pi/10$ between zero and $\pi/2$, considering every possible combination of these. Figure \ref{fig:orientation}b shows how the orientation of the cube changes as it interacts with a soap film. Here, the angles that two of the normal vectors that define the cube make with the $z$ axes are plotted against each other. An empty circle represents the starting orientation of the cube, a dashed line shows how the orientation varies throughout the simulation and the filled circle shows the final orientation of the cube after it has detached from the soap film. The stable orientations can be easily seen in this figure, where the filled circle lies on top of an empty circle with no dashed line to be seen. This is true when the angles $(\alpha_2,\alpha_3)$ are equal to $\left(0,\pi/2\right)$, $\left(\pi/2,0\right)$ and $\left(\pi/2,\pi/2\right)$, which are all representations of the \emph{flat} stable orientation. It is also the case when $(\alpha_2,\alpha_3)=\left(\pi/4,\pi/4\right)$ and $\left(\pi/2,\pi/4\right)$, which are two representations of the cube in the \emph{diagonal} orientation. We also include the result for the \emph{rotated} orientation that we previously discussed, which is shown at $\left(\pi/2-\tan^{-1}\left(1/\sqrt{2}\right),\pi/2-\tan^{-1}\left(1/\sqrt{2}\right)\right)$ here. All other initial orientations are unstable. In all these cases, after an initial perturbation just after attachment, the soap film acts to rotate the cube towards the \emph{flat} orientation. This is shown by the large collection of filled circles at the three corners that represent this orientation in figure \ref{fig:orientation}b. It is clear from this figure that unless the cube is in either the \emph{rotated} or \emph{diagonal} meta-stable orientations, then it is highly likely to be reoriented to the flat orientation as it falls through the centre of a soap film.

\subsection{Variation of initial position}
	\label{sec:position}

So far we have consider what happens when we drop a super-quadric object centrally in a bamboo foam contained in a cylindrical tube. Let us now vary the initial position of the object in the cylinder and inspect how its radial position varies as it interacts with the soap film. Recall that the cylindrical tube has radius $r_c=4$ and is positioned so that its vertical centreline intersects the origin of the Cartesian coordinate system we use. We now vary the initial radial distance, $r_{xy}$, of the centre of the object from the centre-line of the cylinder by varying the value of $x_0$, the initial $x$ coordinate of the object's centre. 

\begin{figure}
    \centering
\includegraphics[width=0.8\textwidth]{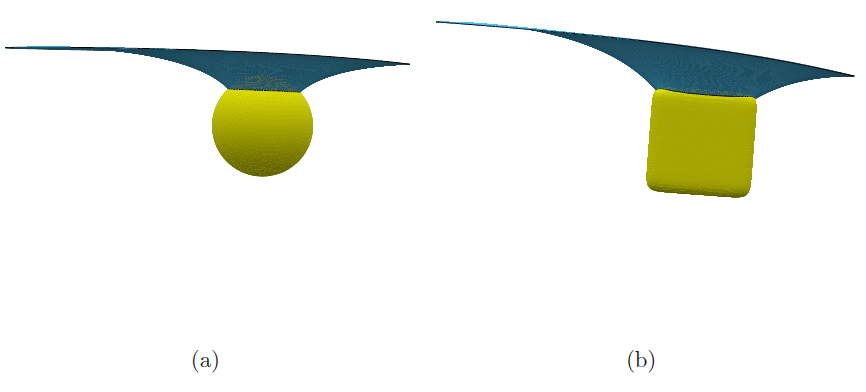}
		\caption{Snapshots of simulations where (a) a sphere and (b) an initially flat cube fall through a soap film from an off-centre position in the cylinder, causing non-symmetric deformation to the soap film.}
			\label{fig:offcentre}
	\end{figure}

As a result of the object being in an off-centre position, the deformation it causes to the foam is asymmetric which affects its resulting motion in the foam. Figure \ref{fig:offcentre} shows how asymmetric the deformation of the soap film becomes when a sphere and a cube are allowed to fall through it from a radial position such that $r_{xy}/r_c=0.25$. 

\begin{figure}
\centering
\includegraphics[width=0.8\textwidth]{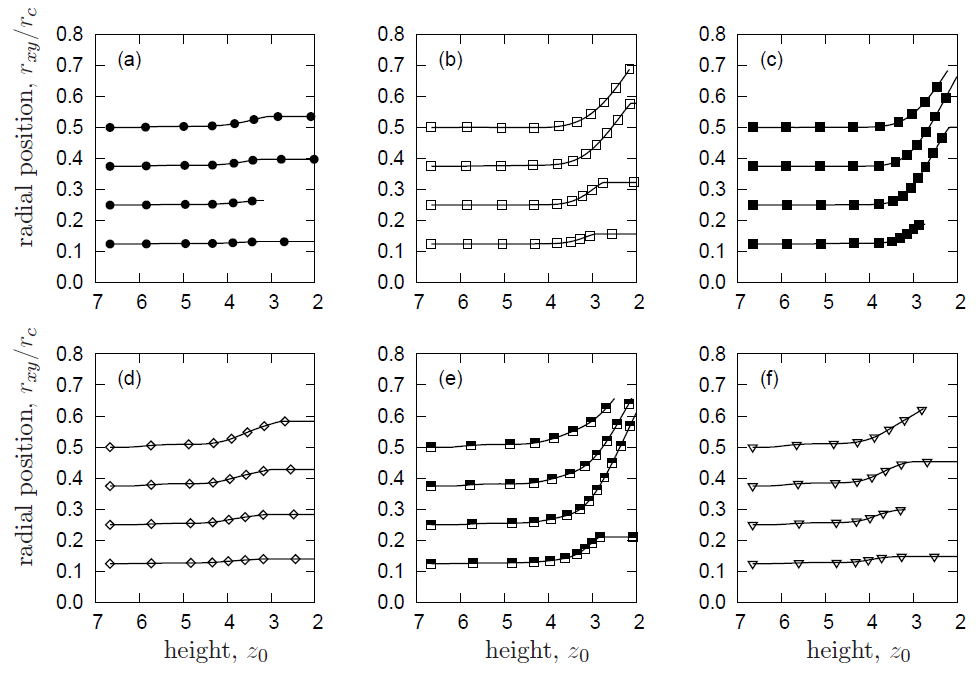}
	\caption{The variation in the radial position, $r_{xy}/r_c=0.25\sqrt{x_0^2+y_0^2}$, of objects versus their height in the foam. Four initial positions are considered: $r_{xy}/r_c=0.125,\,0.25,\,0.375,\,0.5$ for (a) a sphere, (b) a \emph{flat} cube with $\lambda=6$, (c) a \emph{flat} cube with $\lambda=10$, (d) a \emph{diagonal} cube with $\lambda=10$ where the nearest face to the side of the cylinder is vertical, (e) a \emph{diagonal} cube with $\lambda=10$ where the nearest faces to the side of the cylinder are slanted and (f) a \emph{rotated} cube with $\lambda=10$.}
	\label{fig:initial_pos_shapes}
\end{figure}

Figure \ref{fig:initial_pos_shapes} shows how the radial position of the object changes as it interacts with a soap film starting from different initial positions. When a sphere is allowed to fall through a horizontal soap film from an off-centre position in the cylinder, the effects of the asymmetry of the deformation to the film are weak (see figure \ref{fig:initial_pos_shapes}a). There are small increases in the radial distance of the sphere from the centre of the cylinder and these increase slightly with increasing initial radial distance. If the sphere fell through a long bamboo foam containing many bubbles, it would fall towards the wall of the cylinder, and eventually the foam would lose control over its motion. 

Figures \ref{fig:initial_pos_shapes}b and \ref{fig:initial_pos_shapes}c show how the radial position of flat cubes, with $\lambda=6$ and $\lambda=10$ respectively, varies with their height in the foam. It is clear here that the effects of positioning the cube off-centre in the cylinder are stronger than for a sphere. The radial distance of the cube from the centre of the cylinder deviates more here, especially just before it detaches from the soap film, which causes the cube to fall towards the wall of the cylinder. We also note that this effect is stronger for the largest value of $\lambda$, where the deformation caused to the soap film is greater. An indication of why wall effects are more prevalent for a cube compared to a sphere is given in figure \ref{fig:offcentre}; it shows that in particular the angles at which the soap film contacts the cube from the left and right are different. On the side of the cube that is nearest to the cylinder wall (left in figure \ref{fig:offcentre}), the soap film makes a smaller angle with the horizontal than on the other side of the cube. As a result, there is a resultant network force that drags the cube further towards the nearest wall. We also note that the unbalanced deformation of the soap film also tilts the cube slightly, which indicates that the \emph{flat} orientation is no longer stable when the cube is positioned off-centre in the cylinder. The deformation is more symmetric for the sphere, explaining why it deviates less from a vertical path compared to the cube.

Figures \ref{fig:initial_pos_shapes}d and \ref{fig:initial_pos_shapes}e show that a cube in two different \emph{diagonal} orientations also move towards the nearest wall when positioned off-centre in the cylinder. Two different representations of the \emph{diagonal} orientation are considered: one in which the cube has a vertical face closest to the nearest wall and the other where the cube has two slanted faces nearest to the wall. It is clear from figure \ref{fig:initial_pos_shapes}d that the effect the asymmetric deformation has on the motion of the cube is weaker for the first case than for the second orientation shown in figure \ref{fig:initial_pos_shapes}e. When the cube has two slanted faces nearest to the wall, the thin part of the soap film comes into contact with the rounded edge separating them. The angle the soap film makes with the horizontal plane at this stage is close to zero, therefore it pulls the object towards this nearest wall with a large network force. This is similar to what happens in figure \ref{fig:offcentre}bfor the flat cube. The effect is weaker for the diagonal cube that has a vertical face close to the wall, as the thin part of the soap film slips easily along this surface. This is also the case for the rotated cube (see figure \ref{fig:initial_pos_shapes}f), where the attraction towards the wall is of a similar nature to that experienced by a cube in the diagonal orientation considered in figure \ref{fig:initial_pos_shapes}d. This again demonstrates that the soap film navigates the surface of the rotated cube without deforming as much, especially leading up to detachment, where the forces are at their largest.

\section{Conclusions}
	\label{sec:conclusions}

We have presented results of 3D quasi-static simulations in which a light solid object defined by a single super-quadric equation falls under gravity through a horizontal soap film of an ordered bamboo foam. We discussed in detail how the soap film deforms as it attaches to the object and prior to when it detaches from the object. The influence the soap film has over the final position and orientation of the solid object has also been discussed in detail. In particular, we investigated how important the initial shape, orientation and position of the object is to its final position and orientation.

It is clear from this work that the shape of the object is pivotal in how the soap film deforms and the resulting forces the foam exerts on the object. We varied the shape of the object by varying the parameter $\lambda$ between $\lambda=2$ (a sphere) and $\lambda=20$ (a cube with rounded edges and corners), keeping its radius $r_s$ fixed. We found that both the network and pressure drag exerted on the object by the film increase with $\lambda$. This implies that a cube causes the soap film to deform much more than a sphere. This is particularly true in the build up to when the soap film detaches from the object, where it becomes much more stretched than it does around a sphere. In general, it takes the soap film more iterations to navigate surfaces with very high curvature such as the rounded edges of our cubes, especially when those edges are adjacent to flat surfaces that are perpendicular to the direction of the motion of the object.

Another important factor that determines how a film interacts with a solid object is the orientation of the solid object as it comes into contact with it. We considered many orientations of the cube and investigated the deformation to the soap film and the forces it exerts on the object. For a cube dropped centrally through the film, the \emph{flat} orientation was the only stable orientation found, while the \emph{diagonal} and \emph{rotated} orientations were meta-stable. These orientations can be visualised by considering a cube balancing on one face, edge or corner respectively. In all three cases, the deformation to the soap film remains symmetric around the centre-line of the cylindrical container, meaning that the network and pressure torques exerted by the foam remain negligible throughout. The deformation of the soap film is clearly different for these three orientations, in particular during attachment and detachment and this has been be explained using geometrical arguments. For other initial orientations, the symmetry of the soap film deformation is broken, and therefore a non-zero torque is exerted by the foam. This is such that a cube not initially in one of the three aforementioned orientations is likely to be realigned by the horizontal soap film towards the \emph{flat} orientation. The major contribution to this realignment occurs when the forces are at their greatest, that is as the soap film reaches the upper half of the object and before detachment.   

We also investigated the effect on the above results of changing the bubble size. For this, we varied both the radius and height of the cylindrical container and repeated many of the simulations described. This did not affect the magnitude of the network and pressure forces exerted by the film on the falling object or the overall interaction that occurred between the film and the object, and therefore we did not present these results here.  

Finally we investigated the effect of positioning the falling object off-centre in the cylindrical container. This also broke the symmetry of the soap film deformation, resulting in an object being pushed towards the nearest container wall as it fell through the foam. Here, the soap film acts to tip the object towards the wall. This effect was seen to be strongest for a flat cube, as the deformation is exaggerated for this case, and weakest for a sphere or a rotated cube for which a soap film deforms less before detachment.

We therefore conclude that a bamboo foam may provide a good tool for precisely reorienting an object that falls through it from a central position in the cylindrical container, and that for a cube the resulting orientation is highly likely to be the flat orientation. However, a bamboo foam does not provide good control over the motion of objects when they are positioned off-centre in the cylinder. So to improve the control a foam has over the motion of non-spherical objects, we would need to consider more complicated ordered foams such as the staircase, twisted staircase or double staircase foams \cite{Davies12}.

Future work may include an investigation of how the interaction between a soap film and a solid object varies for other shapes such as ellipsoids, as well as the surface properties of the object and the wetting film covering it. For the latter, the contact angle between the soap film and the falling object would need to be varied, introducing the possibility of including frictional forces into our model.
 
\section*{Acknowledgements}

We thank K. Brakke for providing and supporting the Surface Evolver. Funding from the Coleg Cymraeg Cenedlaethol and the MATRIXASSAY H2020-MSCA-RISE-2014 project number 644175 is gratefully acknowledged.


\end{document}